\documentclass[twocolumn,prb,showpacs,superscriptaddress,showkeys,preprintnumbers,amsmath,amssymb]{revtex4}

\usepackage{graphicx}
\usepackage{dcolumn}
\usepackage{bm}
\usepackage{multirow}
\usepackage{subscript}
\usepackage{nicefrac}

\begin{document}


\title{Magnetic structure and magnon dispersion in LaSrFeO$_4$}

\author{N. Qureshi}

\email[Corresponding author. Electronic
address:~]{qureshi@ph2.uni-koeln.de} \affiliation{$II$.
Physikalisches Institut, Universit\"{a}t zu K\"{o}ln,
Z\"{u}lpicher Strasse 77, D-50937 K\"{o}ln, Germany}

\author{H. Ulbrich}

\affiliation{$II$. Physikalisches Institut, Universit\"{a}t zu
K\"{o}ln, Z\"{u}lpicher Strasse 77, D-50937 K\"{o}ln, Germany}

\author{Y. Sidis}

\affiliation{Laboratoire L\'{e}on Brillouin, C.E.A./C.N.R.S., F-91191 Gif-sur-Yvette Cedex, France}

\author{A. Cousson}

\affiliation{Laboratoire L\'{e}on Brillouin, C.E.A./C.N.R.S., F-91191 Gif-sur-Yvette Cedex, France}

\author{M. Braden}

\email{braden@ph2.uni-koeln.de}

\affiliation{$II$. Physikalisches Institut, Universit\"{a}t zu
K\"{o}ln, Z\"{u}lpicher Strasse 77, D-50937 K\"{o}ln, Germany}

\date{\today}

\begin{abstract}

We present elastic and inelastic neutron scattering data on
LaSrFeO$_4$. We confirm the known magnetic structure with the
magnetic moments lying in the tetragonal basal plane, but
contrarily to previous reports our macroscopic and neutron
diffraction data do not reveal any additional magnetic phase
transition connected to a spin reorientation or to a
redistribution of two irreducible presentations. Our inelastic
neutron scattering data reveals the magnon dispersion along the
main-symmetry directions \mbox{[0 $\xi$ 0]} and \mbox{[$\xi$
-$\xi$ 0]}. The dispersion can be explained within linear
spin-wave theory yielding an antiferromagnetic nearest-neighbour
interaction parameter $J_{1}=7.4(1)$ meV and a next-nearest
neighbour interaction parameter $J_{2}=0.4(1)$ meV. The dispersion
is gapped with the out-of-plane anisotropy gap found at
$\Delta_{out}=5.26(2)$ meV, while evidence is present that the
in-plane anisotropy gap lies at lower energies, where it cannot be
determined due to limited instrument resolution.

\end{abstract}


\maketitle
\section{Introduction}
\label{sec:Introduction}

\begin{figure}
\includegraphics[width=0.45\textwidth]{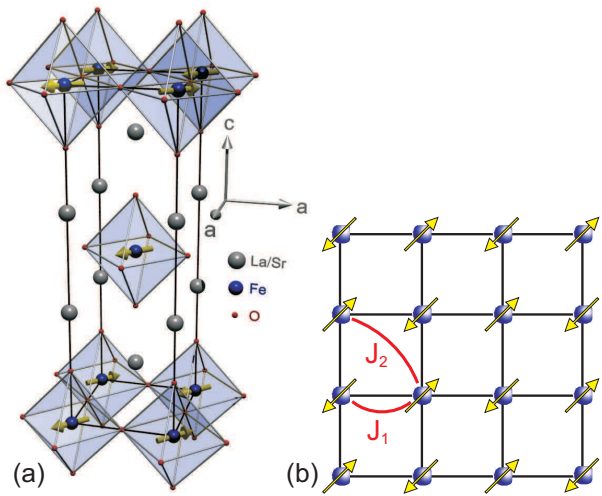}
\caption{\label{fig:structure} (Color online) (a) Visualization of
the crystal and magnetic structure of LaSrFeO$_4$. (b) Tetragonal
basal plane showing only the magnetic Fe ions in order to declare
the coupling constants $J_{1}$ and $J_{2}$.}
\end{figure}

Transition-metal oxides of the Ruddlesden-Popper series
$R_{n+1}M_n$O$_{3n+1}$~(Ref.~\onlinecite{rud1957}) exhibit a large
variety of interesting physical properties such as charge, spin
and orbital ordering, which are intimately coupled and may lead to
fascinating phenomena like the colossal magnetoresistance in
LaMnO$_3$~(Ref.~\onlinecite{jin1994}) ($n=\infty$, '113'
structure). The $n=1$ compound LaSr$M$O$_4$ ('214' structure), a
two-dimensional analog, reveals a single-layered perovskite
structure of the K$_2$NiF$_4$ type [space group $I4/mmm$,
Fig.~\ref{fig:structure}(a)], where the O ions octahedrally
coordinate the $M$ ions to build perfect $M$O$_2$-square planes
[Fig.~\ref{fig:structure}(b)]. While in the '113' compounds these
planes are vertically connected to form a three dimensional
magnetic network, they are separated and shifted along
$[\frac{a}{2} \frac{b}{2} 0]$ in the '214' compounds, which
reduces their electronic dimensionality and renders these systems
ideal for studying their orbital and magnetic correlations in a
less complex environment. For $M$=(Mn, Fe, Co, Ni, Cu) all 214
systems are known to be charge-transfer insulators with an
antiferromagnetic ground
state.\cite{kaw1988,reu2003,sen2005,sou1980,yam1989,vak1987,rod1991,yam1992}
The magnetic structures have been reported to exhibit collinear
spin arrangements, where the nearest neighbour spins are coupled
antiferromagnetically and the next-nearest neighbour spins are
coupled ferromagnetically. However, La$_2$CuO$_4$
(Ref.~\onlinecite{rod1991,yam1992}) and La$_2$NiO$_4$
(Ref.~\onlinecite{vak1987}) exhibit slightly canted
antiferromagnetic structures. LaSrFeO$_4$ orders magnetically at
$T_N$=380~K (Ref.~\onlinecite{sou1980}) and two further magnetic
phase transitions were reported as the susceptibility shows
anomalies at 90 K and 30 K (Ref.\onlinecite{jun2005}). These
magnetic phase transitions are thought to originate from a
redistribution of two collinear
representations.\cite{jun2005}\newline Magnetic excitations in
layered transition metal oxides have attracted considerable
interest in the context of both the high-temperature
superconductors and the manganates exhibiting colossal
magnetoresistance.\cite{ima1998} Rather intense studies on
nickelates, manganates and cobaltates with the K$_2$NiF$_4$ (214)
structure have established the spin-wave dispersion for pure and
doped materials. In the pure materials there is a clear relation
between the orbital occupation and the magnetic interaction
parameters, which may result in unusual excitations like in-gap
modes.\cite{sen2003} Upon doping almost all of these layered
materials exhibit some type of charge ordering closely coupled to
a more complex magnetic order. Most famous examples are the stripe
order in some cuprates and in the nickelates\cite{ima1998} and
also the CE-type order in half-doped manganates.\cite{zal2000}
Recently it was shown that also doped La$_{2-x}$Sr$_{x}$CoO$_4$
(Ref.~\onlinecite{cwi2009}) and La$_{1-x}$Sr$_{1+x}$MnO$_4$
(Ref.~\onlinecite{ulb2011}) exhibit an incommensurate magnetic
ordering closely resembling the nickelate and cuprate stripe
phases when the Sr content deviates from half-doping so that
stripe order can be considered as a general phenomenon in cuprate
and non-cuprate transition-metal oxides.\cite{ulb2012} Magnetic
excitations in these complex ordered materials give a direct
insight to the microscopic origin of these phases. For example, in
La$_{0.5}$Sr$_{1.5}$MnO$_4$ one may easily associate the dominant
magnetic interaction with an orbital ordering.\cite{sen2006} In
comparison to the rather rich literature of manganates, nickelates
and cuprates, there is no knowledge about the magnon dispersion in
LaSrFeO$_4$. We have performed an extensive study of macroscopic
measurements, X-ray and neutron diffraction as well as inelastic
neutron scattering on LaSrFeO$_4$ single crystal and powder
samples in order to address the question of eventual
spin-reorientation phase transitions and to deduce the coupling
constants between nearest
 and next-nearest neighbors within linear spin-wave
theory.

\section{Experimental}
\label{sec:Experimental}

The sample preparation has been carried out similar to reported
techniques.\cite{oka1987,ban2003} Powder samples of LaSrFeO$_4$
have been prepared by mixing La$_2$O$_3$, SrCO$_3$ and Fe$_2$O$_3$
in the stoichiometric ratio and sintering at 1200$^\circ$C for
\mbox{100 h}. Diffraction patterns were taken on a Siemens D5000
X-ray powder diffractometer in order to confirm the correct phase
formation (space group $I4/mmm$) and the absence of parasitic
phases. Furthermore, the lattice constants were deduced at those
temperatures used in the neutron study due to the higher precision
of the powder method.\newline Large single crystals of LaSrFeO$_4$
have been grown by the floating-zone method. Therefore,
LaSrFeO$_4$ powder was pressed into a cylindrical rod of 60 mm
length and 8 mm diameter and sintered at 1300$^\circ$C for 20 h.
The crystal has been grown in a floating-zone furnace (Crystal
Systems Incorporated) equipped with four halogen lamps (1000 W).
The feed and seed rods were rotated in opposite directions at
about 10 rpm, while the molten zone was vertically moved at a
growth speed of 3 mm/h. This procedure has been performed under a
pressure of 4 bars in argon atmosphere. Suitable single crystals
for X-ray diffraction have been obtained by milling larger pieces
in a ball mill for serveral hours. The characterization at the
X-ray single crystal diffractometer Bruker Apex D8 validated the
successful crystal growth. The magnon dispersion has been
investigated at the thermal and cold neutron triple-axis
spectrometers 2T and 4F.2 at the Laboratoire L\'eon Brillouin
(LLB) using a large single crystal of 3.33 g weight, whose single
crystal state was verified at a Laue diffractometer. For energy
transfers above 20 meV inelastic data has been recorded on the 2T
spectrometer, which was used with a pyrolytic graphite (PG)
monochromator and a PG analyzer. The final neutron energy was
fixed at either $E_f=34.9$ meV, $E_f=14.7$ meV, or $E_f=8.04$ meV.
The 4F.2 spectrometer was used with a PG double monochromator and
PG analyzer. A cooled Be filter was used to suppress higher
harmonics. The final neutron energy was fixed at $E_f=4.98$ meV.
\newline The nuclear and magnetic structure determination has been
carried out at the neutron single-crystal diffractometer 5C2 (LLB)
situated at the hot source of the Orph\'ee reactor. For the
elastic measurements a smaller single crystal of 39 mg has been
used. A wavelength of 0.83~\AA~ has been employed supplied by the
(220) reflection of a Cu monochromator. The N\'eel temperature was
derived at the 3T.1 spectrometer (LLB) using a furnace.
Magnetization data was obtained by a commercial superconducting
quantum interference device (SQUID) and a vibrating sample
magnetometer (VSM). Electric resistivity has been measured by the
standard four-contact method.

\section{Results and discussion}
\label{sec:results}

\subsection{Macroscopic properties}
\label{sec:magnetization}

\begin{figure}
\includegraphics[width=0.45\textwidth]{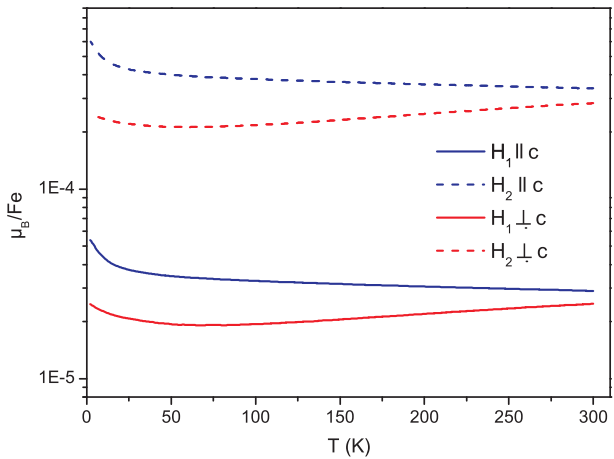}
\caption{\label{fig:squid} (Color online) Magnetization given in
Bohr magnetons per Fe atom as a function of temperature for two
different applied magnetic fields parallel and perpendicular to
the $c$ axis ($H_1$=100 Oe, $H_2$=1000 Oe). No hints for a
magnetic phase transition can be observed between 1.8 K and 300
K.}
\end{figure}

\begin{figure}
\includegraphics[width=0.45\textwidth]{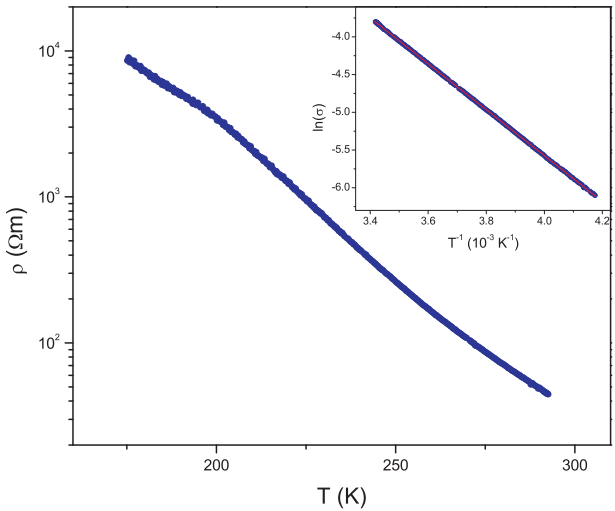}
\caption{\label{fig:rho} (Color online) Specific resistivity of
LaSrFeO$_4$ as a function of temperature. An Arrhenius fit
($\ln{\sigma}=\ln{\sigma_0}-\frac{E}{2kT}$) to the
high-temperature data yields a band gap of $\Delta$=0.525(1) eV.}
\end{figure}

The magnetization data obtained from the SQUID do not show any
additional magnetic phase transitions between 1.8 K and 300 K
(Fig.~\ref{fig:squid}). An additional measurement in a VSM with an
oven did not reveal any signature of $T_N$ in the measured range
from 300 K to 800 K. Such a behaviour is characteristic for the
layered magnetism in 214 compounds, where the ordering temperature
has only been unambiguously determined by neutron diffraction
experiments.\cite{sen2008} The fact that the Fe magnetic moments
exhibit two-dimensional correlations well above $T_N$ renders it
impossible to detect this transition macroscopically. The
transition from two-dimensional to three-dimensional magnetic
order can however be seen via neutron diffraction.
Fig.~\ref{fig:rho} shows the specific resistivity as a function of
temperature. No reliable data could be obtained below 175 K due to
the high values of $\rho$. The high-temperature data has been
plotted in a $\ln(\sigma)$-$\frac{1}{T}$ plot to which an
Arrhenius function ($\ln{\sigma}=\ln{\sigma_0}-\frac{E}{2kT}$) has
been fitted (shown in inset). From the linear behavior a band gap
of $\Delta$=0.525(1) eV can be deduced. At 200 K a kink is visible
in the specific resistivity whose origin is not yet clear to us.

\subsection{Nuclear structure}
\label{sec:nuclear}

\begin{figure}
\includegraphics[width=0.45\textwidth]{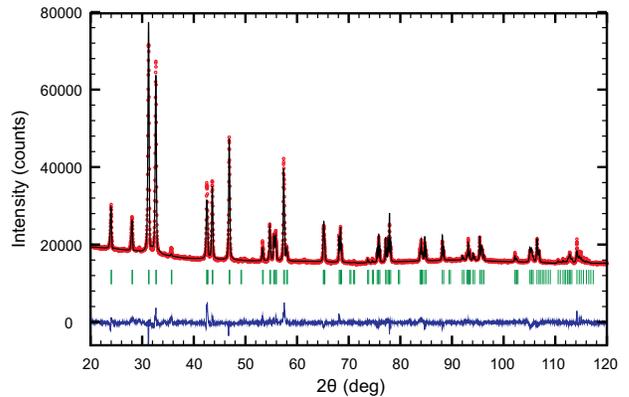}
\caption{\label{fig:powder} (Color online) X-ray powder
diffraction pattern of LaSrFeO$_4$ taken at room temperature. Raw
data are depicted by (red) dots, while the calculated pattern and
the difference line are represented by solid (black) and by solid
(blue) lines, respectively. (Green) vertical bars indicate the
position of Bragg reflections.}
\end{figure}

The investigation of the powder samples by X-ray diffraction
confirmed the reported crystal structure. All powder diffraction
patterns were analyzed using the \textsc{FullProf}
program.\cite{fullprof} Fig.~\ref{fig:powder} depicts the powder
pattern recorded at room temperature. The calculated pattern
[(black) solid line] agrees very well with the observed pattern
[(red dots)] and no parasitic peaks can be observed. Additional
diffraction patterns were recorded at 120 K, 50 K and 10 K. The
lattice parameters have been deduced and are listed in
Tab.~\ref{tab:structure}.
For the nuclear structure investigation at the neutron single
crystal diffractometer a total number of 739 independent
reflections has been collected at each temperature. The integrated
intensities were corrected for absorption applying the
transmission factor integral $\exp[-\mu (\tau_{in}+\tau_{out})]$
by using subroutines of the \textsc{Cambridge Crystallographic
Subroutine Library}\cite{ccsl} ($\tau_{in}$ and $\tau_{out}$
represent the path lengths of the beam inside the crystal before
and after the diffraction process, $\mu$ is the linear absorption
coefficient, which is 0.146 cm$^{-1}$ for LaSrFeO$_4$). The
nuclear structure refinement included the $z$ value of the atomic
positions of La/Sr and O2, the anisotropic temperature factors of
all ions (respecting symmetry restrictions according to
Ref.~\onlinecite{pet1966}), the occupation of the O1 and O2 site
as well as the extinction parameters according to an empiric
ShelX-like model.\cite{lar1970} All refined structural parameters
are shown in Tab.~\ref{tab:structure}. The atomic positions show
almost no significant dependence on the temperature, while the
lattice constants and anisotropic displacement parameters (ADP)
expectedly decrease with decreasing temperature. The only
exception are the $U_{33}$ parameters, which increase for all
species when reducing the temperature from 50 K to 10 K. This
results in a much more anisotropic atomic displacement at 10 K
with more out-of-plane motion. One may realize that some of the
ADPs are stronger than it might be expected from the phononic
contributions. It has already been pointed out in earlier studies
on La$_{2-x}$Sr$_{x}$CuO$_4$ (Ref.~\onlinecite{bra2001}) and
La$_{1+x}$Sr$_{1-x}$MnO$_4$ (Ref.~\onlinecite{sen2005}) that the
intrinsic disorder due to the occupation of the same site by La
and Sr causes a non-zero force on the oxygen ions at the mean
atomic positions derived by diffraction experiments. Due to the
La/Sr-O bonds being perpendicular to the Fe-O bonds the disorder
will affect the displacement of the O ions mainly perpendicular to
the Fe-O bonds, i.e. the $U_{33}$ parameter of O1 and the $U_{11}$
parameter of O2 are most affected. The refinement indeed yields
pronounced enhancement of these parameters. From the neutron data
an eventual oxygen deficiency might be deduced. Taking into
account the refinement with the best agreement factors one can
calculate the stoichiometry of the investigated compound to be
LaSrFeO$_{3.92(6)}$ yielding a slight oxygen deficiency.
\newline

\begin{table*}
\caption{\label{tab:structure}Nuclear structure parameters within
the $I4/mmm$ space group at different temperatures. The Wyckoff
sites are La/Sr 4$e$ (00$z$), Fe 2$a$ (000), O1 4$c$
(0$\frac{1}{2}$0) and O2 4$e$ (00$z$). The results of the neutron
single crystal diffraction experiment were completed by measuring
the lattice
constants using X-ray powder diffraction. 
For La/Sr, Fe and O2 $U_{22}$ is constrained by
symmetry to $U_{11}$.}
\begin{ruledtabular}
\begin{tabular}{cccccc}
       & T (K) & 10 K & 50 K & 120 K & RT\\ \hline 
       & $a$ (\AA) & 3.8709(1)   &  3.8713(1)  &  3.8726(1) & 3.8744(1)\\ 
       & $c$ (\AA) & 12.6837(4)  &  12.6848(4) &  12.6931(4)& 12.7134(3)\\ 
La/Sr  & $z$  & 0.3585(1) & 0.3589(1) & 0.3589(1) &  0.3587(1)\\ 
       & $U_{11}$ (\AA$^2$)  & 0.0032(6) & 0.0056(5) & 0.0067(5) & 0.0107(5)\\ 
       & $U_{33}$ (\AA$^2$)  & 0.007(1) & 0.0045(4) & 0.0044(4) & 0.0089(5)\\ 
Fe     & $U_{11}$ (\AA$^2$)  & 0.0017(7) & 0.0036(4) & 0.0047(4) & 0.0069(6)\\ 
       & $U_{33}$ (\AA$^2$)  & 0.013(1) & 0.0099(5) & 0.0121(5) & 0.0186(7)\\ 
O1     & occ (\%)            & 99(2)    & 102(2)    & 102(2) &
99(2)\\ 
       & $U_{11}$ (\AA$^2$)  & 0.0050(9) & 0.009(1) & 0.007(2) & 0.011(1)\\ 
       & $U_{22}$ (\AA$^2$)  & 0.0034(9) & 0.003(2) & 0.006(2) & 0.0074(9)\\ 
       & $U_{33}$ (\AA$^2$)  & 0.012(1) & 0.0085(4) & 0.0097(5) & 0.0175(9)\\ 
O2     & occ (\%)            & 97(2)    & 97(2)     & 97(2)    &
96(2)\\ 
       & $z$           & 0.1694(2) & 0.1686(1) & 0.1689(1) & 0.1692(2)\\ 
       & $U_{11}$ (\AA$^2$)  & 0.0160(9) & 0.0160(4) & 0.0179(7) & 0.0230(8)\\ 
       & $U_{33}$ (\AA$^2$)  & 0.011(1) & 0.0079(7) & 0.0077(6)  & 0.0124(7)\\ 
       \hline
       & $R_F$ (\%)    & 2.65 & 2.84 & 3.17 & 2.78\\ 
       & $\chi^2$      & 0.33 & 2.35 & 2.05 & 4.60\\ 
\end{tabular}
\end{ruledtabular}
\end{table*}

\subsection{Magnetic structure}
\label{sec:magnetic}

\begin{figure}
\includegraphics[width=0.48\textwidth]{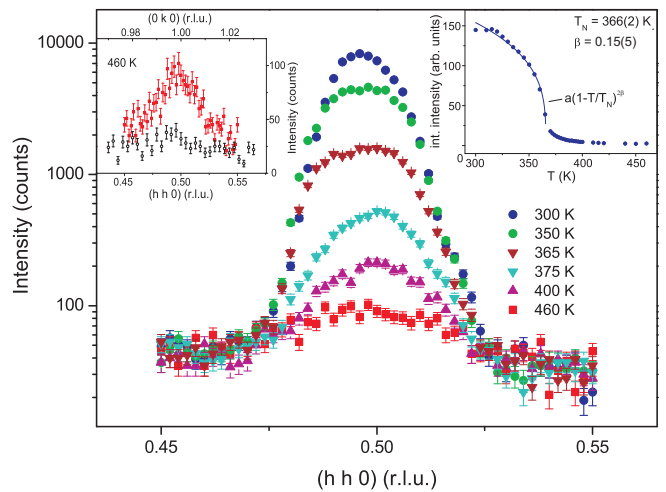}
\caption{\label{fig:TN} (Color online) Longitudinal
$\mathbf{q}$-scan across the magnetic $(\frac{1}{2} \frac{1}{2}
0)$ reflection (the intensity is plotted on a log-scale). The
inset at the upper right shows the integrated intensity as a
function of temperature. A power-law fit yields a $T_N$ of 366(2)
K. The inset at the upper left depicts the same longitudinal
$\mathbf{q}$-scan at 460 K [(red) filled squares] together with a
longitudinal $\mathbf{q}$-scan across the forbidden (010)
reflection [(black) open circles] documenting that no $\lambda/2$
contamination is present.}
\end{figure}

We found strong half-indexed magnetic Bragg peaks confirming the
known propagation vector $\mathbf{q}_1=(\frac{1}{2} \frac{1}{2}
0)$. The intensity of the magnetic $(\frac{1}{2} \frac{1}{2} 0)$
reflection was measured as a function of temperature and is shown
in Fig.~\ref{fig:TN}. A power-law fit to the integrated intensity
data yields a N\'eel temperature of 366(2) K and a critical
exponent $\beta$=0.15(5) (upper right inset). However, an exact
determination of $T_N$ is hardly possible as significant intensity
due to strong quasielastic scattering can be observed well above
the transition temperature e.g. at 400 K or 460 K. By scanning
across the forbidden (010) reflection an eventual $\lambda/2$
contamination can be ruled out (upper left inset).

\begin{figure}
\includegraphics[width=0.49\textwidth]{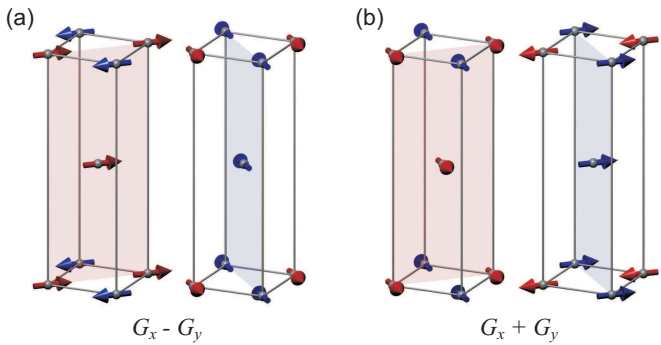}
\caption{\label{fig:domains} (Color online) Magnetic G-type
ordering (according to standard Wollan-Koehler
notation\cite{wol1955}) in LaSrFeO$_4$ only showing the magnetic
Fe ions and their spins for the two $\mathbf{q}$-domains. (a)
shows the irreducible representation labelled $G_x-G_y$ which has
the magnetic moments perpendicular to $\mathbf{q}_1$ (left) and
$\mathbf{q}_2$ (right). (b) shows the irreducible representation
labelled $G_x+G_y$ with the magnetic moments parallel to
$\mathbf{q}_1$ (left) and $\mathbf{q}_2$ (right).}
\end{figure}

As two additional magnetic phase transitions might be expected at
90~K and 30~K~(Ref.~\onlinecite{jun2005}) the magnetic structures
have been investigated at 120~K, 50~K and 10~K. For the magnetic
structure refinement a total number of 185 independent reflections
has been recorded at each temperature point, where the integrated
intensities have been corrected for absorption. Representation
analysis has been used to derive symmetry adapted spin
configurations which were then refined to the respective data.
Three irreducible representations are compatible with the space
group $I4/mmm$ yielding collinear spin configurations with the
moments parallel to the $c$ axis, parallel to $\mathbf{q}$ or
perpendicular to $\mathbf{q}$, the last two being of orthorhombic
symmetry.\footnote{As the magnetic structure possesses
orthorhombic symmetry it might be expected that the crystal
structure reacts to the onset of magnetism by an orthorhombic
distortion. Therefore, additional high resolution powder
diffraction data have been collected using synchrotron radiation
(B2, HASYLAB). However, no peak splitting could be observed within
the instrumental resolution.} Due to the fact that
$\mathbf{q}_2=(-\frac{1}{2}\frac{1}{2}0)$ is a possible
propagation vector as well each of the irreducible representations
with the basis vectors in the {\it a-b} plane will exhibit two
magnetic orientations. The domains are connected to each other by
the symmetry operator (y,-x,z) which has been lost during the
transition into the magnetically ordered state. The relevant spin
configurations used in previous analyses\cite{sou1980,jun2005} are
shown in Fig.~\ref{fig:domains}.


\begin{figure}
\includegraphics[width=0.45\textwidth]{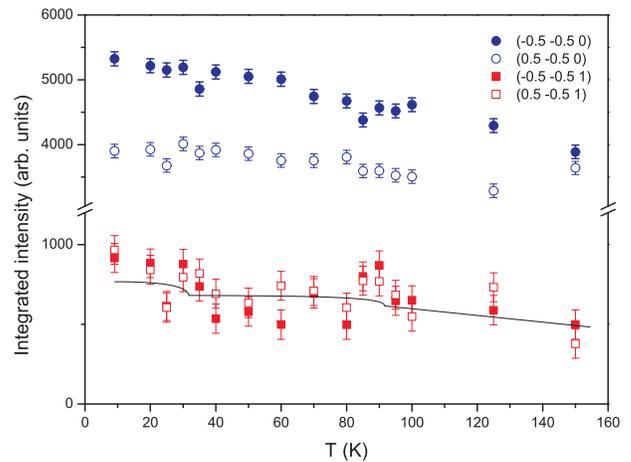}
\caption{\label{fig:nojumps} (Color online) Integrated intensity
of characteristic magnetic Bragg reflections probing either domain
$\mathbf{q}_1$ (positive $h$ values) or domain $\mathbf{q}_2$
(negative $h$ values) as a function of temperature. The solid line
represents the size of the jumps reported in
Ref.~\onlinecite{jun2005}.}
\end{figure}

The data could well be described by the model $G_x - G_y$, where
the size and the direction (angle $\phi$ between the moment and
the $a$ axis) of the magnetic moments in the basal plane as well
as the percental distribution between the two magnetic domains
were refined. The results are listed in Tab.~\ref{tab:magnetic}
for all investigated temperatures. In Refs.~\onlinecite{sou1980}
and~\onlinecite{jun2005} the authors claim that their sample
exhibits an inhomogeneous distribution of the two collinear
representations $G_x-G_y$ and $G_x+G_y$ accounting for 92\% and
8\% of the sample volume. Furthermore, the intensity jumps of
characteristic magnetic Bragg reflections at the transition
temperatures 30 K and 90 K were attributed to a change in the
relative distribution of the representations. We have followed the
integrated intensity of characteristic magnetic Bragg reflections
as a function of temperature across the two lower magnetic phase
transitions and could not observe any significant jumps
(Fig.~\ref{fig:nojumps}). Although the statistics seem to be
limited in comparison to the size of the jumps at least for the
(0.5 0.5 1) reflection, one can state that the scattering from
both domains does not exhibit contrary behavior in dependence of
temperature ruling out a redistribution of domain population. We
have applied the proposed inhomogeneous distribution of two
representations to our data, however, no significant contribution
of $G_x+G_y$ is present (see Tab.~\ref{tab:magnetic}).

\begin{table}
\caption{\label{tab:magnetic}Results of the magnetic structure
refinement. }
\begin{ruledtabular}
\begin{tabular}{cccc}
 T (K) & 10 K & 50 K & 120 K  \\ \hline
$m_{Fe}$ ($\mu$\textsubscript{B}) & 4.96(1) & 5.38(1) & 5.09(1) \\
$\phi$ (deg) & 44.0(8) & 44.9(8) & 46.7(8) \\
domain $\mathbf{q}_1$ (\%) & 46.5(4) & 47.3(4) & 47.7(5) \\
domain $\mathbf{q}_2$ (\%) & 53.5(4) & 57.7(4) & 52.3(5) \\
$G_x-G_y$ (\%) & 98(2) & 97(2) & 98(2) \\
$G_x+G_y$ (\%) &  2(2) & 3(2)  & 2(2) \\
\hline
 $R_F$ (\%)    & 5.14 & 4.43 & 4.76  \\
 $\chi^2$      & 1.78 & 1.75 & 1.95  \\
\end{tabular}
\end{ruledtabular}
\end{table}

\subsection{Magnon dispersion}
\label{sec:magnon}

\begin{figure}
\includegraphics[width=0.45\textwidth]{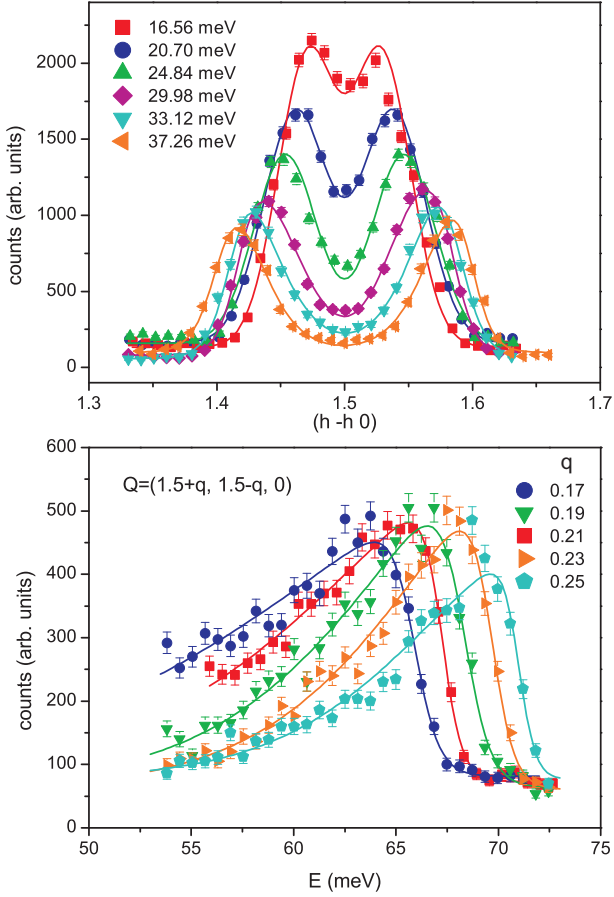}
\caption{\label{fig:scans} (Color online) Exemplary scans showing
how the values of $S(q,\omega)$ have been obtained. Constant-E
scans have been fitted with two symmetric Gauss functions (upper
plot), while constant-$\mathbf{Q}$ scans have been fitted with
asymmetric double sigmoids (lower plot).}
\end{figure}

The magnon dispersion has been investigated at \mbox{10 K} along
the two main-symmetry directions \mbox{[0 $\xi$ 0]} and
\mbox{$[\xi$ -$\xi$ 0]}. Depending on the orientation of the
resolution ellipsoid with respect to the dispersion branch
constant-$\mathbf{Q}$ or constant-E scans have been performed. The
excitation signals have been fitted with two symmetrical Gauss
functions (constant-$\mathbf{Q}$ scans) or an asymmetric
double-sigmoid\footnote{The asymmetric double-sigmoid function $I
= bgr +
A\cdot[1+\exp(-\frac{x-x_c+w_1/2}{w_2})]^{-1}\cdot[1-(1+\exp(-\frac{x-x_c-w_1/2}{w_3}))^{-1}]$
makes it possible to describe asymmetric peak profiles with
unequal parameters $w_2$ and $w_3$.} (constant-E scans) in order
to account for the strong asymmetry at high energy transfers. An
asymmetry has been applied to the symmetric Gauss functions for
the constant-$\mathbf{Q}$ scans at higher energy transfer.
Exemplary scans are shown in Fig.~\ref{fig:scans} documenting the
data analysis.

\begin{figure}
\includegraphics[width=0.48\textwidth]{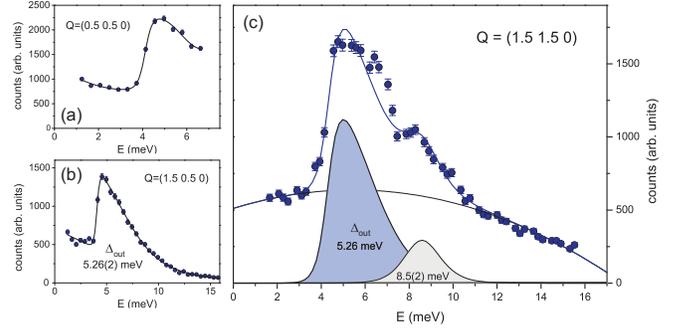}
\caption{\label{fig:gaps} (Color online) Constant-$\mathbf{Q}$
scans at the magnetic zone centers (a) (0.5 0.5 0), (b) (1.5 0.5
0) and (c) (1.5 1.5 0). The excitations have been fitted by
asymmetric double sigmoid functions on a polynomial background.
The out-of-plane anisotropy gap amounts to 5.26(2) meV as obtained
by a fit to the data in panel (b). In panel (c) the signal at
8.5(2) meV is presumably phononic as the same scan at lower
$\mathbf{Q}$ values, see panel (b), is featureless at this energy.
$\Delta_{out}$ has been held constant in the fit to the data shown
in panel (c). The fit curve in panel (a) serves as a guide to the
eye. The considerable scattered intensity below $\Delta_{out}$ is
a hint for the lower-lying in-plane excitation.}
\end{figure}

According to the magnetic structure, where the magnetic moments
are lying in the tetragonal basal plane, two gapped excitations
have to be expected. The lower one should be connected to the
amount of energy needed to turn a spin out of its ordered position
within the basal plane whereas the higher one results from turning
a spin out of the plane. Constant-$\mathbf{Q}$ scans have been
performed at different Brillouin zone centers in order to derive
the size of the respective spin gaps. In
Fig.~\ref{fig:gaps}(a)-(c) a clear signal can be observed at
5.26(2) meV [the value has been obtained from an asymmetric
double-sigmoid fit to the scan at $\mathbf{Q}$=(1.5 0.5 0)], which
we identify as the higher-lying out-of-plane gap $\Delta_{out}$.
In Fig.~\ref{fig:gaps}(c) an additional signal appears at 8.5(2)
meV which, however, is not present in the lower zone center scans
and therefore rather phononic than magnetic. It can be seen
especially in Fig.~\ref{fig:gaps}(a) and (b) that the scattered
intensity is not reduced to the background below $\Delta_{out}$.
Bearing in mind the energy resolution of 0.23 meV as obtained from
the FWHM of the elastic line we conclude that the scattered
intensity at low energy originates from the in-plane fluctuation
of the magnetic moments. Due to the finite size of the resolution
ellipsoid and its inclination in $S(q,\omega)$ space signals from
steep dispersion branches become very broad as can be seen in
Fig.~\ref{fig:gaps}(b), where considerable scattered intensity is
observed well above 10 meV. For this reason we expect the in-plane
fluctuations to be gapless.

Within linear spin wave theory we used a Hamiltonian of a
Heisenberg antiferromagnet with isotropic nearest ($J_1$) and
next-nearest neighbor ($J_2$) interactions [see
Fig.~\ref{fig:structure}(b)] as well as an effective magnetic
anisotropy field $H_A$ along the $z$ axis\cite{mar1971}
\begin{align}
H
&=\sum_{\mathbf{m,r}}J_1(\mathbf{r})\mathbf{S_m}\cdot\mathbf{S_{m+r}}+\sum_{\mathbf{m,R}}J_{2}(\mathbf{R})\mathbf{S_m}\cdot\mathbf{S_{m+R}}\notag \\
&+\sum_{\mathbf{n,R}}J_{2}(\mathbf{R})\mathbf{S_n}\cdot\mathbf{S_{n+R}}+g\mu_BH_A\left(
\sum_m S^z_{\mathbf{m}}-\sum_n
S^z_\mathbf{n}\right).\notag \\
\end{align}

Here the magnetic lattice has been divided into two identical
sublattices $m$ and $n$ where each of them only contains parallel
spins. $\mathbf{r}$ is a connection vector between magnetic
moments of the interpenetrating antiferromagnetically coupled
($J_1$) sublattices with respective positions $\mathbf{m}$ and
$\mathbf{n}$, while $\mathbf{R}$ denotes a connection vector
between two ferromagnetically coupled magnetic moments of the same
sublattice ($J_2$). Each spin pair contributes only once to the
sum. The diagonalization of the Hamiltonian\cite{mar1971} leads to
the dispersion relation for this particular crystal structure:

\begin{align}
\hbar\omega_q&=\{[4SJ_{1}-4SJ_{2}[1-\cos{(2\pi q_x)}\cos{(2\pi q_y)}]\notag \\
&+g\mu_BH_A]^2-(2SJ_{1})^2[\cos{(2\pi q_x)}+\cos{(2\pi
q_y)}]^2\}^\frac{1}{2}. \label{eq:dispersion}
\end{align}

\begin{figure}
\includegraphics[width=0.48\textwidth]{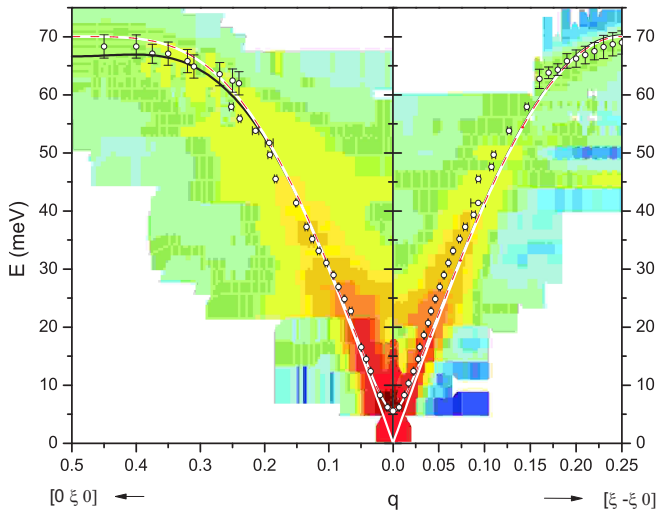}
\caption{\label{fig:disp} (Color online) Two-dimensional
reconstruction of $S(q,\omega)$ along the two symmetry directions
\mbox{[0 $\xi$ 0]} and \mbox{[$\xi -\xi 0$]} from all
constant-$\mathbf{Q}$ and constant-E scans as exemplarily shown in
Fig.~\ref{fig:scans}. Dots mark the peak center as obtained from
the fits to the raw data. The straight white line is a fit of the
dispersion relation (Eq.~\ref{eq:dispersion}) to both data sets
simultaneously.}
\end{figure}

We have fitted the dispersion relation simultaneously to spin
waves propagating along \mbox{[0 $\xi$ 0]} and \mbox{[$\xi$ -$\xi$
0]}. With an expected $S=2.5$ we obtain $J_{1}=7.4(1)$ meV,
$J_{2}=0.4(1)$ meV and $\mu_B H_{A,out}=0.097(2)$ T (note that a
factor 2 has been added to the $J$ values for a correct comparison
with Refs.~\onlinecite{lar2005,nak1993,bab2010,col2001} due to a
different definition of the sums in the Hamiltonian). The
agreement with the experimental data is fairly well, the
dispersion curve is depicted as a black solid line in
Fig.~\ref{fig:disp}. Setting $J_2$=0 yields $J_1$=6.99(1) and
$\mu_B H_{A,out}=0.102(2)$ T and the agreement is comparable
[(red) dashed line in Fig.~\ref{fig:disp}]. While along
\mbox{[$\xi$ -$\xi$ 0]} the dispersion is practically unchanged,
it goes to higher energy values at the zone-boundary for the
propagation along \mbox{[0 $\xi$ 0]}, however, staying within the
error bars of the data points. The fact that $J_2$ is not
essentially needed to describe the dispersion makes it possible to
apply the spin-wave dispersion reported in
Ref.~\onlinecite{thu1982}, which has been derived for the
isostructural K$_2$FeF$_4$ structure. This spin Hamiltonian for
Fe$^{2+}$ in a tetragonally distorted cubic crystal field only
contains a nearest-neighbour exchange parameter, but considers two
non-degenerate spin-wave dispersion branches, which - in a
semiclassical picture - correspond to elliptical precessions of
the spins with the long axis of the ellipse either in or
perpendicular to the layer.\cite{thu1982} The spin-wave dispersion
is given in Eq.~\ref{eq:thudisp} for the larger orthorhombic cell

\begin{align}
\hbar\omega_q&=4S|J|[(1+A)^2\notag\\
&-(\cos{[\pi(q_x+q_y)]}\cos{[\pi(q_x-q_y)]}\pm B)^2]^{\frac{1}{2}}
\label{eq:thudisp}
\end{align}

with $A=(D-3E)/(4|J|)$ and $B=(D+E)/(4|J|)$. $D$ is a parameter
describing the uniaxial anisotropy and $E$ adds an in-layer
anisotropy. Fitting Eq.~\ref{eq:thudisp} with $E=0$ to both data
sets simultaneously yields the values $J$=7.00(1) meV and
$D$=0.0409(6) meV. The two non-degenerate branches are depicted as
white solid lines in Fig.~\ref{fig:disp}. The upper branch
coincides exactly with the formalism in Eq.~\ref{eq:dispersion}
(only nearest-neighbour interaction) yielding the same coupling
constant within the error bars. The lower branch is gapless as
predicted by $E$=0 and the reason for the non-zero intensity below
$\Delta_{out}$ in Fig.~\ref{fig:gaps}.


An examination of the energy gaps as a function of temperature
yielded no differences between 10 K and 100 K.

\section{Conclusion}
\label{sec:discussion}

\begin{figure}
\includegraphics[width=0.45\textwidth]{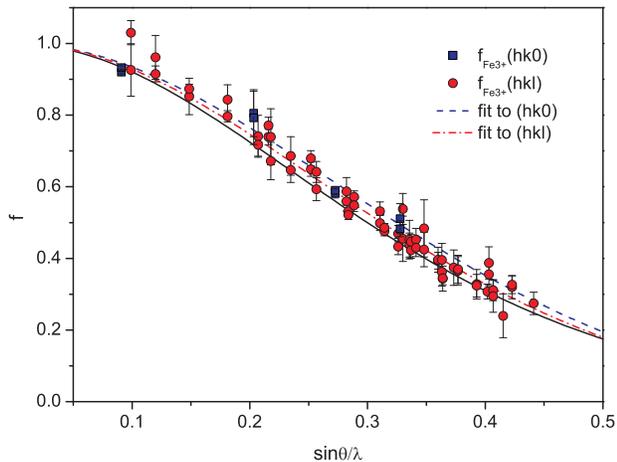}
\caption{\label{fig:magff} (Color online) Observed magnetic Fe
form factor for (hk0) [(blue) squares] and (hkl) [(red) dots]
magnetic reflections. The (black) solid line depicts the tabulated
analytical approximation of the Fe$^{3+}$ form fator. The data
have been fitted by varying the tabulated values while requiring
f(0)=1 [(red) dash-dotted line is the fit to the (hkl) data,
(blue) dashed line is the fit to the (hk0) data].}
\end{figure}

We have conducted a comprehensive study on the single-layered
perovskite LaSrFeO$_4$. Our X-ray powder and single crystal
diffraction as well as Laue diffraction yield high sample quality,
while our magnetization data differ from previously published
work. Detailed investigation of the nuclear structure by neutron
diffraction on a single crystal reveals that the intrinsic
disorder on the La/Sr site leads to a stronger atomic displacement
of the O1 and O2 ions in LaSrFeO$_4$ in analogy to
La$_{1+x}$Sr$_{1-x}$CuO$_4$ (Ref.~\onlinecite{bra2001}) and
La$_{1+x}$Sr$_{1-x}$MnO$_4$ (Ref.~\onlinecite{sen2005}). The main
results of our study concern the magnetic structure and the magnon
dispersion of this compound, which we analyzed by neutron
diffraction and inelastic neutron scattering. We have addressed
the open question concerning the magnetic phase transitions at 90
K and 30 K. Our SQUID data did not yield any hint for additional
magnetic phase transitions and based on our neutron diffraction
data we are able to say that no spin reorientation or
domain/representation redistribution is present. Possible
discrepancies between the data of different studies might be the
exact amount of oxygen as these systems are known to exhibit
oxygen deficiency. From the nuclear structure refinement we can
deduce the oxygen deficiency to be $y=0.08(6)$ in LaSrFeO$_{4-y}$.
The large number of measured reflections allows an analysis of the
magnetic form factor. Therefore, the observed magnetic structure
factors were divided by the exponential part of the calculated
magnetic structure factor and the ordered moment. The resulting
observed magnetic form factor has been derived for (hk0) and (hkl)
reflections in order to gain information about the in-plane and
out-of-plane atomic magnetization density distribution in
LaSrFeO$_4$. In Fig.~\ref{fig:magff} the observed magnetic form
factors for both kinds of magnetic reflections are depicted
showing a tendency towards weaker decrease with increasing
$\sin(\theta)/\lambda$ in comparison to the tabulated analytical
approximation of the Fe$^{3+}$ magnetic form factor, which would
imply a more localized atomic magnetization density distribution.
However, due to the limited number of (hk0) reflections and the
size of the error bars no significant anisotropy can be deduced.

In addition, we have analyzed the magnon dispersion along the two
symmetry directions \mbox{[0 $\xi$ 0]} and \mbox{[$\xi$ -$\xi$
0]}. Within linear spin wave theory we can describe both branches
with the nearest neighbor and next-nearest neighbor interaction
$J_1=7.4(1)$ meV and $J_2=0.4(1)$ meV, respectively ($S=2.5$).
These values are more than a factor 2 larger than those in the
isostructural undoped LaSrMnO$_4$ with $S=2$
(Ref.~\onlinecite{lar2005}), which can be attributed to the fact
that $e_g$-$e_g$ superexchange contributes only little for the
case of the Mn orbital arrangement in LaSrMnO$_4$. Our data
indicate two non-degenerate spin-wave dispersion branches. A clear
signal at 5.26(2) meV was identified as the out-of-plane
anisotropy gap $\Delta_{out}$. The non-zero scattered intensity at
lower energy transfers is explained by the lower-lying anisotropy
gap, which was then analyzed by using the formalism described in
Ref.~\onlinecite{thu1982}. With a nearest-neighbour interaction
$J$=7.00(1) meV and the anisotropy parameters $D$=0.0409(6) and
$E$=0 a good agreement between the upper dispersion branch and the
experimental data has been achieved, while the lower branch goes
down to zero-energy transfer at the magnetic zone center. The spin
wave dispersion in La$_2$NiO$_4$ is also well described by a
nearest neighbor interaction only,\cite{nak1993} but with $J=31$
meV it is a factor of 4 stronger than in LaSrFeO$_4$ indicating
higher hybridization in La$_2$NiO$_4$. Although a more involved
Hamiltonian has been used for the description of the spin dynamics
in La$_2$CoO$_4$ (with high-spin $S$=$\frac{3}{2}$ Co$^{2+}$)
including three coupling constants and corrections for spin-orbit
coupling, ligand and exchange fields,\cite{bab2010} the resulting
coupling constants are of the same order as the ones presented
here. Furthermore, the out-of-plane gap and the bandwidth are
quite comparable. In order to compare the single-ion anisotropy of
the involved species within one model we have used
Eq.~\ref{eq:dispersion} together with the $J$ values given in
Refs.~\onlinecite{lar2005,nak1993,bab2010,col2001} (up to
next-nearest neighbour exchange) to calculate the anisotropy
parameter $\mu_B H_{A,out}$. The approximate values of the
dispersion at the zone center and zone boundary were taken from
plots within Refs.~\onlinecite{lar2005,nak1993,bab2010,col2001}.
The comparison of $J$ and $H_{A,out}$ for the different compounds
is shown in Tab.~\ref{tab:JHcomp}.

\begin{table}
\caption{\label{tab:JHcomp} Comparison of the exchange coupling
and anisotropy parameters of LaSrFeO$_4$ with other 214 compounds
[LaSrMnO$_4$ (Ref.~\onlinecite{lar2005}), La$_2$CoO$_4$
(Ref.~\onlinecite{bab2010}), La$_2$NiO$_4$
(Ref.~\onlinecite{nak1993}), La$_2$CuO$_4$
(Ref.~\onlinecite{col2001})].}
\begin{ruledtabular}
\begin{tabular}{cccccc}
 & Mn & Fe & Co & Ni & Cu \\ \hline
$S$        &  2     & 2.5    &  1.5    & 1   & 0.5 \\
$J_1$ (meV)& 3.4(3) & 7.4(1) & 9.69(2) & 31  & 104(4)  \\
$J_2$ (meV)& 0.4(1) & 0.4(1) & 0.43(1) & 0   & -18(3)  \\
$\mu_B H_{A,out}$ (T)& 0.65   & 0.097(2) & 0.67 & 0.52 & 0\\
\hline
\end{tabular}
\end{ruledtabular}
\end{table}

One can see that the single-ion anisotropy of the Fe$^{3+}$ in
LaSrFeO$_4$ is significantly smaller than in the other 214
compounds (except for La$_2$CuO$_4$), which is expected due to the
close to zero orbital moment and therefore very weak spin-orbit
coupling.

\begin{acknowledgments}
This work was supported by the Deutsche Forschungsgemeinschaft
through the Sonderforschungsbereich 608.

\end{acknowledgments}

\bibliography{literatur}

\end{document}